\documentclass[runningheads]{llncs}
\usepackage{graphicx}
\usepackage{listings} 
\begin{document}
\title{A Picture Is Worth a Thousand Words: Exploring Diagram and Video-Based OOP Exercises to Counter LLM Over-Reliance}


%
%
\titlerunning{A Picture Is Worth a Thousand Words}
%


\author{Bruno Pereira Cipriano\inst{1} \and
Pedro Alves\inst{1} \and
Paul Denny\inst{2}}

\authorrunning{B. Pereira Cipriano et al.}
%
\institute{Lusófona University, COPELABS, Campo Grande, 376, Lisbon, Portugal
\email{\{bcipriano,pedro.alves\}@ulusofona.pt}
\and
The University of Auckland, Auckland, New Zealand\\
\email{paul@cs.auckland.ac.nz}}
\maketitle              
\begin{abstract}
Much research has highlighted the impressive capabilities of large language models (LLMs), like GPT and Bard, for solving introductory programming exercises. Recent work has shown that LLMs can effectively solve a range of more complex object-oriented programming (OOP) exercises with text-based specifications. This raises concerns about academic integrity, as students might use these models to complete assignments unethically, neglecting the development of important skills such as program design, problem-solving, and computational thinking. To address this, we propose an innovative approach to formulating OOP tasks using diagrams and videos, as a way to foster problem-solving and deter students from a copy-and-prompt approach in OOP courses. We introduce a novel notation system for specifying OOP assignments, encompassing structural and behavioral requirements, and assess its use in a classroom setting over a semester. Student perceptions of this approach are explored through a survey (n=56). Generally, students responded positively to diagrams and videos, with video-based projects being better received than diagram-based exercises. This notation appears to have several benefits, with students investing more effort in understanding the diagrams and feeling more motivated to engage with the video-based projects. Furthermore, students reported being less inclined to rely on LLM-based code generation tools for these diagram and video-based exercises. Experiments with GPT-4 and Bard’s vision abilities revealed that they currently fall short in interpreting these diagrams to generate accurate code solutions.

\keywords{object-oriented programming \and large language models \and gpt-4 \and bard}
\end{abstract}
\section{Introduction}

The advent of large language models (LLM) and their ability to generate computer code from natural language descriptions has led to robust discussion in the computing education community around the opportunities and challenges they present to educators and students ~\cite{becker2023programming,denny2024CACM}.  In fact, there is differing opinion amongst educators regarding whether to resist and fight the usage of these tools, or to try and find ways to embrace them~\cite{lau2023ban,sheard2024instructor}. While there have been some initial attempts to integrate LLMs into teaching practice at the introductory level~\cite{porter2023learn,kazemitabaar2023studying}, very little is known about the efficacy of these approaches and there is little discussion or consensus about how higher level courses should adapt~\cite{prather2023robots}. 

Given that LLMs are becoming an essential part of industry practice~\cite{barke2023grounded}, it is necessary for educators to explore approaches that promote the acquisition of core computing knowledge and skills alongside authentic use of code-generation tools.  Tasks which are solvable through pure ``copy-and-prompt'' approaches may not be engaging or motivating for students. Very recent research has suggested using image-based exercises that illustrate expected behaviors through diagrammatic descriptions of input/output pairs ~\cite{lau2023ban,denny2023promptly}. The goals of this approach, which has only been explored at the introductory programming level, are twofold: first, students have to make an effort to understand the expected transformation, and, second, they can’t just copy the assignment into ChatGPT, Bard or similar tool.


In this research we propose extending the idea of image-based problem specifications to more complex design-oriented tasks suitable for Object-Oriented Programming (OOP) courses.  In such courses, the focus is not on implementing stand-alone functions but rather on tasking students with designing and implementing multiple classes that maintain mutable states and collaborate with each other to achieve specific objectives.  Our aim is also to develop an approach where problems can be presented in such a way that that their solution is not uniquely specified.  Instead, rather than guiding students towards a single model solution, we want students to have the flexibility to infer their own object-oriented design by performing a domain analysis.  To achieve this, we propose a novel notation for expressing program requirements.  Problems are then presented to students using this notation, rather than as highly descriptive plain text specifications, which we hypothesize will help to guard against probable LLM-abuse as motivated in prior work~\cite{denny2023prompt}.


We conducted an experiment with diagram-based OOP exercises and video-based OOP projects, both utilizing a custom notation, and evaluated it in a classroom setting over the course of a semester.
Our investigation is guided by the following three research questions (RQs):

\textbf{RQ1}: Do students express a preference towards diagram-based exercises compared to exercises with more traditional text-based 
specifications?


\textbf{RQ2}: For larger-scale projects, are video-based specifications motivating for students and do they find them easier for interpreting required behavior?

\textbf{RQ3}: To what extent does the proposed notation for specifying programming tasks discourage students from inappropriate use of LLMs for code-generation?



Recently, both ChatGPT\footnote{ChatGPT received `vision' support in September 25, 2023 (only available for paying subscribers) ~\cite{open-AI-chat-gpt-with-vision}.} and Bard\footnote{Google Bard started supporting image input in July 13, 2023 ~\cite{google-bard-image-input}}  were updated with the capacity to interpret image content (i.e. `vision'). We performed some experiments to determine if these new capabilities could jeopardize our diagramming efforts. Although these were ad-hoc experiments, the relevance to our study made us include them in a brief section of this paper.




This paper makes the following contributions: (1) Presents a novel notation to represent OOP exercises; (2) presents the results of a student survey evaluating diagram-based and video-based exercises, as well as the impact of these new exercise formats on their LLM usage; and, (3) presents the results of several ad-hoc experiences using ChatGPT-4 and Bard's `vision' capabilities to solve diagram-based exercises.

\section{Related work}
\label{section:related_work}



\textbf{Specifying Programming Tasks}
Many programming courses require students to implement certain behaviours described in natural language (e.g. ``Create a function that receives an array of int(s) and returns the sum of all elements in the array''). Students have to interpret the textual description of the problem, devise an algorithm to solve it, and then create computer code that implements the algorithm. This is particularly common for introductory programming exercises 
~\cite{allen2019many}. GPT-based models have demonstrated great capacity for solving such exercises described in natural language~\cite{finnie2022robots,Savelka_2023,Savelka_2023_trhilled,denny2023conversing}. 

\textbf{OOP Courses} Some object-oriented programming exercises are based on Unified Modeling Language (UML) diagrams: teachers first teach students how to interpret UML class diagrams, and then ask them to create the code that implements the classes described in the UML diagram ~\cite{ouh2023chatgpt}. Other educators use text based instructions, with either strongly directed instructions~\cite{Savelka_2023,Savelka_2023_trhilled}, or with less directed instructions which require that students partially decide the object model for themselves ~\cite{cipriano:2024}. GPT-based models have also demonstrated the capability to solve text-based OOP exercises~\cite{cipriano2023gpt3,Savelka_2023,Savelka_2023_trhilled,ouh2023chatgpt,cipriano:2024}.

\textbf{LLM-oriented exercises} Several strands of work have explored ways to limit student reliance on LLMs and foster new skillsets.  Denny et al. proposed Prompt Problems ~\cite{denny2023promptly}, a novel pedagogical approach to the teaching of programming, in which diagrams are displayed to students, who must then derive an effective LLM-prompt to obtain working code. Liffiton et al. describe CodeHelp~\cite{liffiton2023codehelp}, a LLM-powered tool which acts an intermediate between the student and GPT, in order to prevent students from over-relying on the LLM by filtering out generated source code. Students reported positive feedback regarding its usefulness for support while programming. 



\section{Notation: Diagrams and videos}





\subsection{Diagrams}



Our proposed notation aims to address the variety of scenarios that typically occur in OOP, such as functions that change the state of an object, functions that receive multiple objects and associate them with each-other, concepts with common attributes and/or behaviours that should take advantage of the inheritance and polymorphism mechanisms, and so on.

We propose 5 types of diagrams, and explain the relevant notation in the following subsections.  A given assignment may use a combination of several diagram types.

\subsubsection{Algorithmic function diagram}

\begin{figure}
  \centering
  \includegraphics[width=0.4\textwidth]{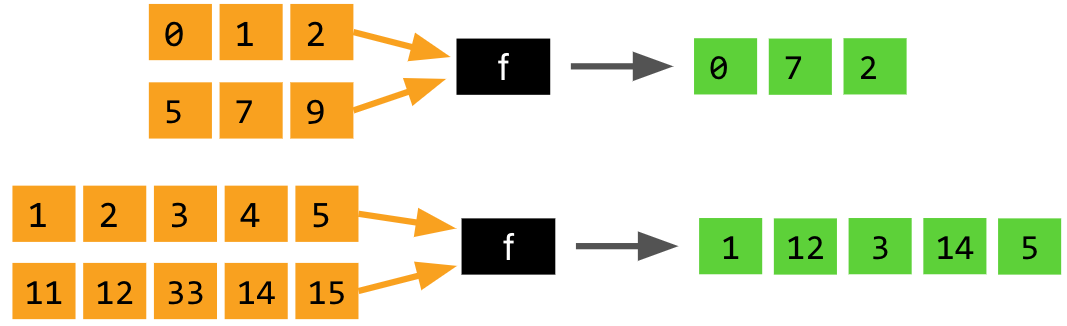}
  \caption{Example algorithmic exercise. Students should understand that the function must create an array with elements from the two input arrays, in alternating fashion.}
  \label{fig:diagram-example-algorithmic}
\end{figure}

These diagrams are used to present students with single non-instance (i.e. static) functions that transform input into output, similar to prior research adapting introductory programming courses ~\cite{denny2023promptly}. See Figure \ref{fig:diagram-example-algorithmic} for an example. 

These diagrams depict the function, with a black box denoting its name, input elements represented using inbound orange arrows, and output elements in green (with outbound black arrows). In cases where inputs or outputs are arrays, lines of small boxes should be utilized, as illustrated in the Figure \ref{fig:diagram-example-algorithmic}.
The function name within the black box should be ``obfuscated,'' so as not to provide clues to the function's purpose (e.g., `f()'). This is recommended as research has shown GPT-based models can generate working solutions from only meaningful function names ~\cite{babe2023studenteval,yeticstiren2023evaluating}.

A minimum of two examples of inputs/outputs should be provided to assist in clarifying the expected behavior. In certain instances, more than two examples may be necessary, particularly to illustrate how the function should behave with invalid inputs or boundary cases.

\subsubsection{State-change function diagram}

These diagrams are similar to the algorithmic function diagrams described in the previous section but are used for non-instance (static) functions that \textbf{change the state of one or more objects} and possibly also return a value. While this may be considered bad practice in imperative programming, where students should avoid implementing functions with side-effects, it is common practice in OOP. Since these functions can have two simultaneous effects, a different notation must be used for changing the objects and returning a value. The former is represented by a dashed arrow while the latter is represented by a solid arrow (in concordance with the algorithmic diagrams that only return values).
Given that the distinction between solid and dashed arrows may not be immediately evident, a caption is provided to highlight this difference for students. The remaining notation is expected to be inferred by students.

The business rules can be directly written in the diagram as a side note but we consider it more interesting to just provide several examples and allow students to infer those rules. For more complex rules, it may not be feasible to describe them through examples - consider using a state transition rules diagram (see Section \ref{section:state-transition-rules}) in those cases.

\begin{figure}
  \centering
  \includegraphics[width=0.4\textwidth]{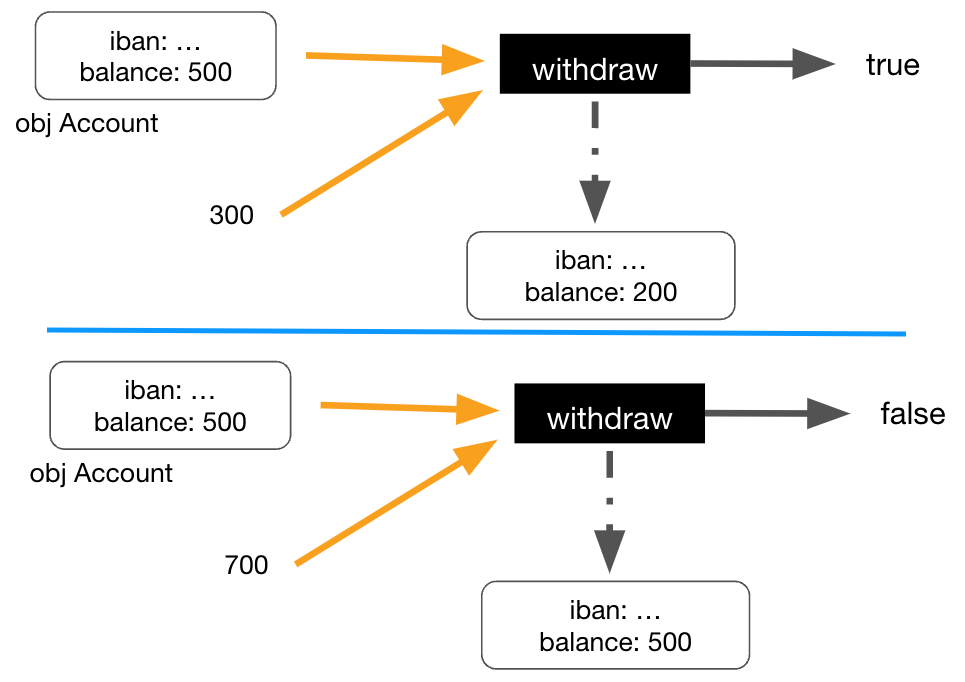}
  \vspace{-5pt}
  \caption{State-change function diagram for the withdraw function. It changes the `Account' object's state and returns true if the operation succeeded, or false otherwise.}
  \label{fig:withdraw-diagram}
\end{figure}

Notice that students are expected to implement not only the function but also the classes and methods that support its behavior.   Imagine you want students to implement the `withdraw' function in Java, as illustrated in Listing~\ref{lst:withdraw_example}. Using our notation, the corresponding diagram is the one shown in Figure~\ref{fig:withdraw-diagram} with the first example showing a successful withdrawal that returns true and reduces the account's balance and the second example showing a failed withdrawal due to insufficient balance.

\begin{lstlisting}[caption={Withdraw function and supporting class, in Java}, label={lst:withdraw_example}, basicstyle=\scriptsize\ttfamily, breaklines=true, breakindent=0pt, frame=single, captionpos=b, float]
class Account {
   public boolean widthdraw(int ammount) {
      if (ammount < this.balance) ...
   }
}

static boolean withdraw(Account account, int ammount) {
   return account.withdraw(ammount);
}
\end{lstlisting}

\subsubsection{Class declaration diagram}

These diagrams provide students with guidelines for implementing predefined classes, attributes, and relationships (e.g., composition). Each class should be depicted within a box, enumerating its attributes (optionally with their types), specifying the necessary constructors, and outlining the expected behavior of fundamental methods like `toString()'. The behavior can be described similarly to algorithmic diagrams, with examples of input/output and corresponding arrows.
It is important to note that the direct representation of relationships between classes is omitted. Specific attributes are linked with an accompanying image, indicating that they are, in fact, objects themselves.

\begin{figure}
  \centering
  \includegraphics[width=0.5\textwidth]{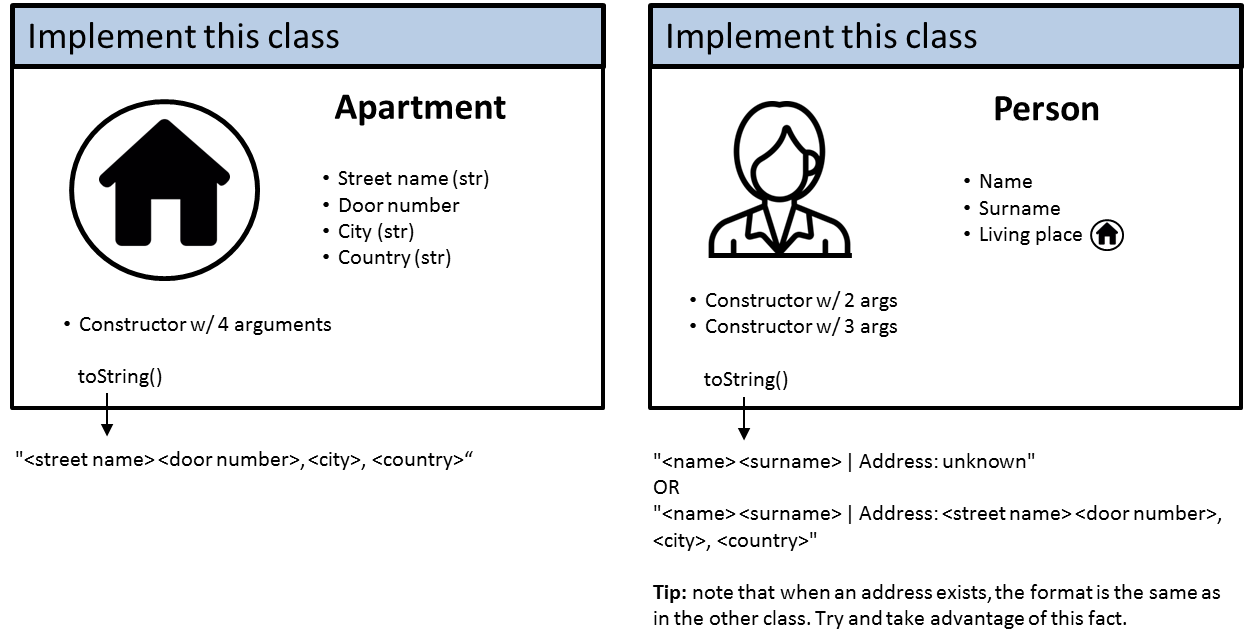}
  \vspace{-10pt}
  \caption{Class declaration exercises. Students should infer that the `Person' class has a composition relationship with the `Apartment' class. Some data-types were omitted, since students should be able to derive them (e.g. name should be a `String').}
  \label{fig:diagram-class-declarations}
\end{figure}

See Figure \ref{fig:diagram-class-declarations} for an example where students had to implement two related classes. In that example, students had to infer that the small house-like figure within the `Person' class diagram represented a composition relationship between the `Person' and `Apartment' classes. Note also the reference to the `toString()' function, with a black arrow which indicates the expected return value, similar to the previously discussed diagrams for algorithmic exercises.


Although this information could be partially represented by UML class diagrams, this proposal has two advantages: (1) it presents both structure (e.g. the class's attributes) and behaviour (e.g. `toString()''s expected return-value); and, (2) the composition relationship is less explicit than it would be in UML.


\subsubsection{Inheritance diagram}

These diagrams are an extension (or adaptation) of the previously mentioned class declaration diagrams, specifically tailored for exercising inheritance relationships between classes (i.e. generalization via inheritance). The notation is exactly the same, however the goal is different. Instead of describing all the classes that the students must implement (as in class diagrams), they only describe the child classes and students must determine the structure and behaviour of the parent class (which is absent from the diagram). After this, they must implement all classes (structure and behavior), both those explicitly outlined in the diagram and classes that are not present in the diagram but have been identified by the students.  

\begin{figure}
  \centering
    \includegraphics[width=0.5\textwidth]{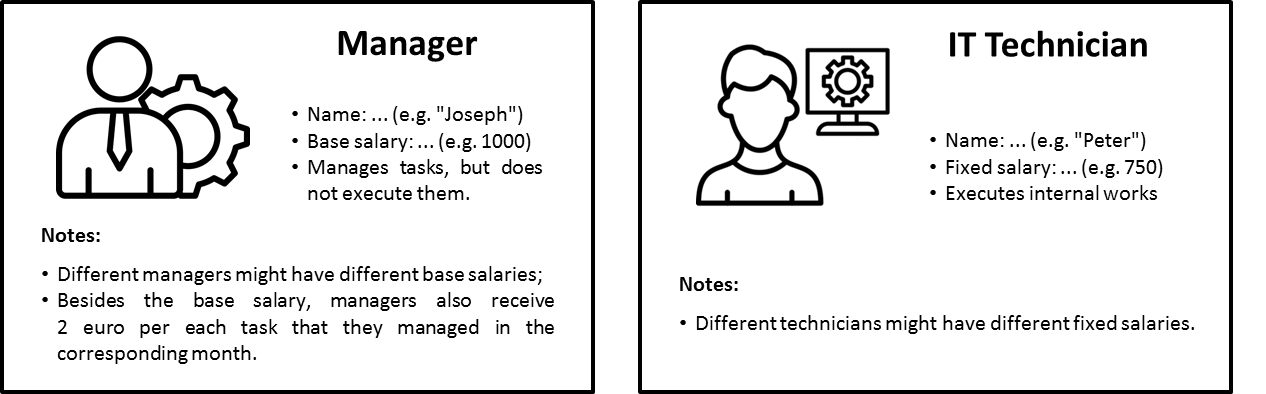}
    \vspace{-10pt}
    \caption{Example inheritance exercise. Students were expected to understand that some common elements exist between the `Manager' and the `IT Technician' and thus create an inheritance relationship with a super-class above those.}
  \label{fig:diagram-example-inheritance}
\end{figure}

Figure \ref{fig:diagram-example-inheritance} presents one of these diagrams, in which two similar concepts are represented: one representing `Managers' and the other representing `IT Technicians'. Students are expected to infer that these two concepts should share a common super-class since they have common attributes (e.g. name, salary) and similar behaviours (e.g. salary calculation).

\subsubsection{State transition rules diagram}
\label{section:state-transition-rules}
\begin{figure}
  \centering
  \includegraphics[width=\columnwidth]{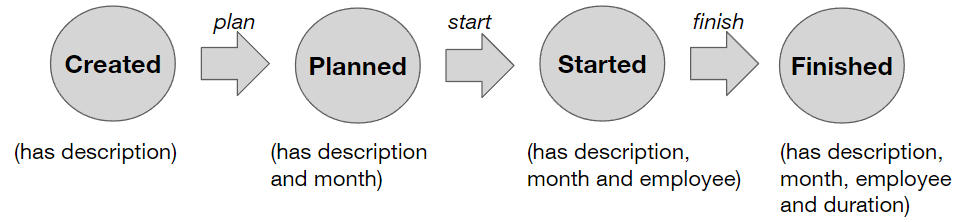}
  \vspace{-10pt}
  \caption{State transition rules. This is a simplified state transition diagram which only display valid transitions.}
  \label{fig:diagram-task-state-transitions}
\end{figure}

These diagrams should be used when the exercise has certain classes which can transition between different internal states, with rules which guide those transitions. In practical terms, these diagrams are simplified state transition diagrams which only display valid transitions. See Figure \ref{fig:diagram-task-state-transitions} for an example. That figure's diagram explains the name of the action which changes the state (e.g. `plan', `start'), as well as what information should exist inside the object in each state. For example, a newly created task only has value for the `description'.

\begin{figure}
  \centering
\includegraphics[width=\columnwidth]{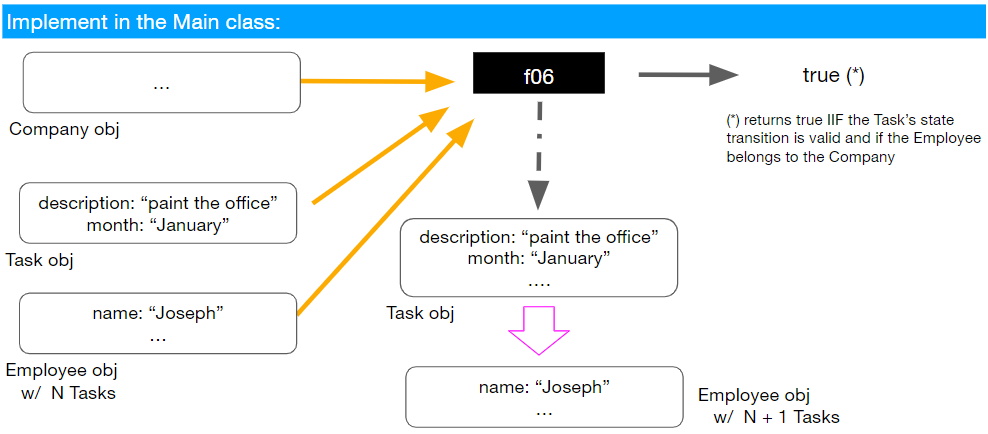}
\vspace{-10pt}
  \caption{Example state-change function exercise. Students were expected to understand that the function would receive 3 objects (a `Company', an `Employee', and a `Task') and associate them if some rules were respected. The rules were described in another diagram which can be seen in Figure \ref{fig:diagram-task-state-transitions}.}
  \label{fig:diagram-example-oop}
\end{figure}

These diagrams are useful for providing complementary information to the other exercises. For example, the state transition diagram of Figure \ref{fig:diagram-task-state-transitions} and the state-change function diagram of Figure \ref{fig:diagram-example-oop} were part of the same assignment and are complementary: for students to correctly implement the `f06()' function, to assign a `Task' to an `Employee', they must understand that the association can only be done if the Task is in the `Planned' state. The diagram explains this connection to the student in the form of a small asterisk near the function's return value.

\subsection{Videos}

While some exercises are easy to describe with a diagram containing 2 or 3 example inputs, or even with 2 or 3 diagrams, in more complex assignments, where there is a significant amount of user interaction or input validation, or projects which span multiple weeks or months, creating diagrams to represent all relevant behaviours and interactions is complex and time consuming.

In such cases, it is more practical to create video demonstrations of the expected behaviours. The videos can include a mixture of behavioural demonstrations and static diagrams. More concretely, we recommend using demonstrations for explaining user interactions, and diagrams for presenting any mandatory protocols which the students must follow. The videos should start with the demonstration of the expected functionality from the user's perspective, and only afterwards go into required implementation details.
A video\footnote{Narrated in Portuguese, but auto-translated captions are available.} with an example OOP course project is available online\footnote{\url{https://www.youtube.com/watch?v=LkyEaAVK6yU}}.





\begin{table}[]
\caption{The quantitative questions used in the survey, organized by Research Question.}
\begin{tabular}{|l|l|l|}
\hline
\textbf{\footnotesize RQ} & \textbf\footnotesize {Q. \#} & \textbf{\footnotesize Question}                                                                                     \\ \hline
\footnotesize 1           & \footnotesize 1                 & \footnotesize \begin{tabular}[c]{@{}l@{}}`It is easier to understand the objective of the exercise when presented in \\the form of a diagram, than when it is presented textually.'                  \end{tabular}                                                                 \\ \hline
\footnotesize 1           & \footnotesize 2                 & \begin{tabular}[c]{@{}l@{}}\footnotesize `When I come across a diagram-based exercise, I tend to think more  carefully \\before writing code, compared to what I do with exercises described textually.'\end{tabular}                 \\ \hline
\footnotesize 1           & \footnotesize 3                 & \footnotesize \begin{tabular}[c]{@{}l@{}}`In general, I prefer exercises based on diagrams over exercises \\described textually.'               \end{tabular}                                                                                                                          \\ \hline
\footnotesize 1           & \footnotesize 4                 & \begin{tabular}[c]{@{}l@{}}\footnotesize `Considering the interpretation of the exercise, the absence of the function \\name in the diagrams ...' \end{tabular}
\\ \hline

\footnotesize 2           & \footnotesize 1                 & \footnotesize \begin{tabular}[c]{@{}l@{}}`It is easier to interpret the project when presented in video format, \\due to the combination of text, images and audio.'\end{tabular}                                                                                                      \\ \hline
\footnotesize 2           & \footnotesize 2                 & \footnotesize \begin{tabular}[c]{@{}l@{}}`Video statements fall short when compared to statements in natural language, \\because it is more difficult to take notes on paper'\end{tabular}                                                                                                   \\ \hline
\footnotesize 2           & \footnotesize 3                 & \footnotesize \begin{tabular}[c]{@{}l@{}}`I feel more motivated to do the project when the statement is in video \\than when it is in natural language.' \end{tabular}                                                                                                                 \\ \hline
\footnotesize 2           & \footnotesize 4                 & \footnotesize \begin{tabular}[c]{@{}l@{}}`I find it easier to develop the project with the video statement than with \\ the traditional model'.  \end{tabular}                                                                                                                         \\ \hline
\footnotesize 3           & \footnotesize 1                 & \footnotesize \begin{tabular}[c]{@{}l@{}}`I am more likely to use GPT/Bard with textual exercise descriptions than \\ with diagrams and/or videos'.\end{tabular}                                                                                                                       \\ \hline
\footnotesize 3           & \footnotesize 2                 & \footnotesize \begin{tabular}[c]{@{}l@{}}`Diagram and/or video exercises effectively prevent abuse of GPT/Bard, \\as they force me to prepare a prompt instead of simply copying the \\ \footnotesize description into GPT/Bard.'\end{tabular} \\ \hline
\footnotesize 3           & \footnotesize 3                 & \footnotesize 
\begin{tabular}[c]{@{}l@{}}\footnotesize ‘I consider diagram and/or video exercises a good step towards making me more \\prepared for a professional future where I have to interact with GPT/Bard.’\end{tabular}         \\ \hline


\end{tabular}
\vspace{-10pt}
\label{tab:table_questions}
\end{table}

\section{Method}

\subsection{Experimental context}


To evaluate the proposed diagrams and videos as formats to present students with OOP tasks, we applied them in the 2023/24 edition of a University course focused on Object-Oriented Design and Programming. In this course, students are exposed to OO analysis and design, as well as Java implementation details. They are also challenged to solve a number of practical programming activities, with the course following a mixed approach with both exercise-based and project-based learning ~\cite{lenfant2023project}, supported by an AAT~\cite{paiva2022automated}.




The diagrammatic notation and videos proposed in this paper were applied to the majority of these practical activities, which were originally described in natural language. Diagrams were used to describe the exercises, while the project was described in video.

\subsection{Evaluation}

To assess the effectiveness of this novel notation, students were administered a structured anonymous questionnaire during the 12th week of the course. At this juncture, 
students had been exposed to multiple diagrams of all types presented in this paper: this includes 9 assignments, each composed of multiple diagrams\footnote{The total number of unique diagrams was 72.}. Also, students were actively engaged in the project.

The questionnaire was composed of 11 quantitative questions, each associated with one RQ. Ten of the questions -- presented in Table \ref{tab:table_questions} -- used a standard 5-point Likert scale (strongly disagree to strongly agree). One question was categorical, with three possible options. Finally, the questionnaire included a qualitative question for students to provide open-response comments.






\section{Results}

The course had 115 enrolled students. Among them, 84 students had some level of participation, submitting at least one diagram-based assignment. For those 84 students, the average number of submitted assignments was 6.25 and the mean was 7 (out of the 9 available assignments). The minimum number of submitted assignments was 1, the maximum and mode were both 10, and 12 students submitted all assignments. A total of 56 results participated in the survey.


\subsection{RQ1: Preference for diagram-based exercises}

\begin{figure}
  \centering
  \includegraphics[width=\columnwidth]{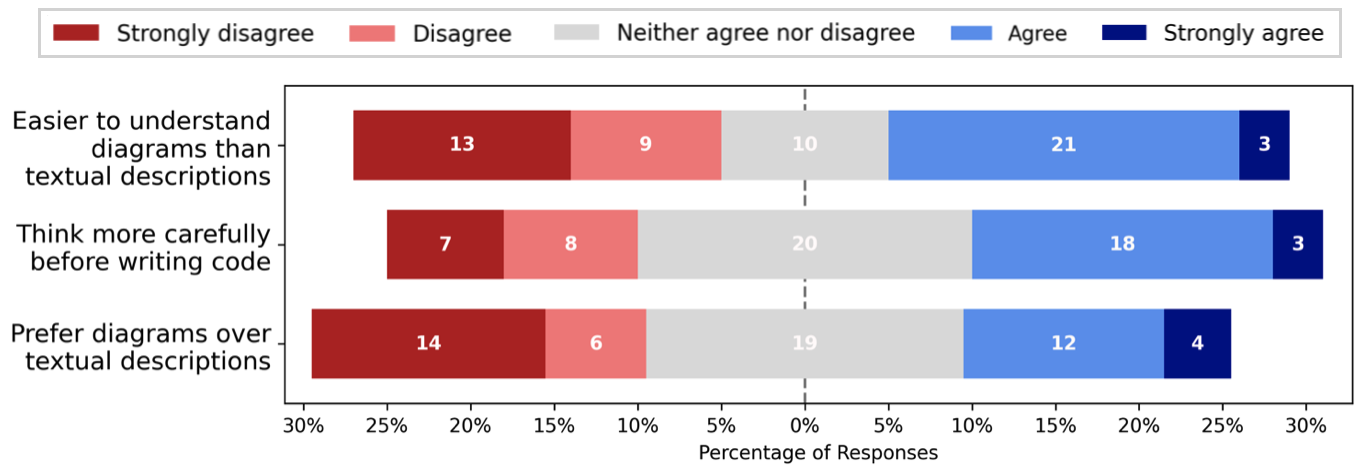}
  \vspace{-10pt}
  \caption{Students opinions with regards to diagram-based exercises when compared to natural language exercises (RQ1).}
  \label{fig:table1}
\end{figure}

This section of the survey was dedicated to understanding students opinions on diagram-based exercises, in comparison to text-based exercises that they are familiar with from previous courses (RQ1). Response distributions to the three quantitative questions associated with RQ1 are presented in Figure \ref{fig:table1}. Although the results are somewhat balanced between the agreement and disagreement sides, in the first two questions, more students agree with the benefits of diagrammatic exercises. However, in the third question, results were skewed to the disagreement side, indicating that most students do not prefer diagram-based exercises over text-based ones.



Finally, we asked students to evaluate the impact of the absence of the function name from the diagrams (Table \ref{tab:table_questions}: Question 4, RQ1). This question was categorical and had 3 possible options: `It causes me some difficulties' (selected by 37 students or 66.07\%), `It doesn't affect me' (14 students or 25.00\%), and, `It causes me a lot of difficulties' (5 students or 8.93\%). These results somewhat surprised us, since we thought that the majority of students would report the lack of function names to have caused them more difficulty. 



\begin{figure}
  \centering
  \includegraphics[width=\columnwidth]{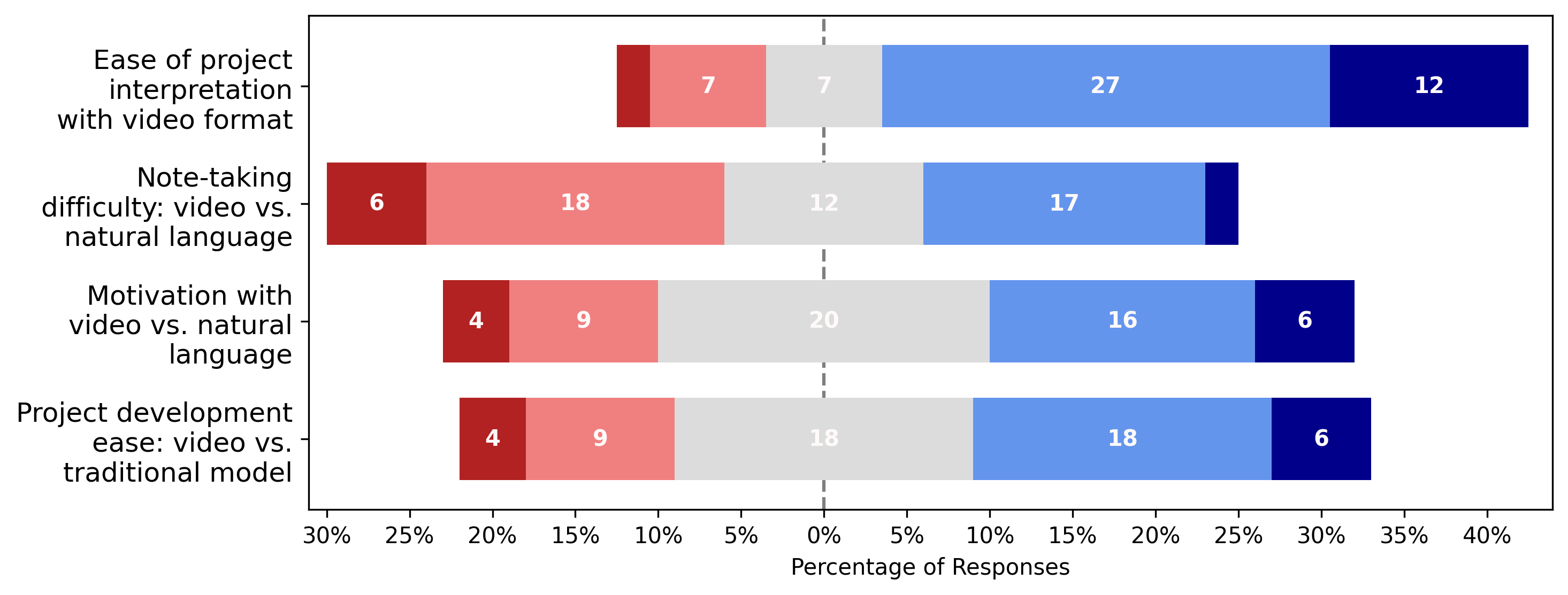}
    \vspace{-10pt}
  \caption{Distribution of student replies to questions related with video-based exercises (RQ2).}
  \label{fig:table3}
\end{figure}



\begin{figure}
  \centering
  \includegraphics[width=\columnwidth]{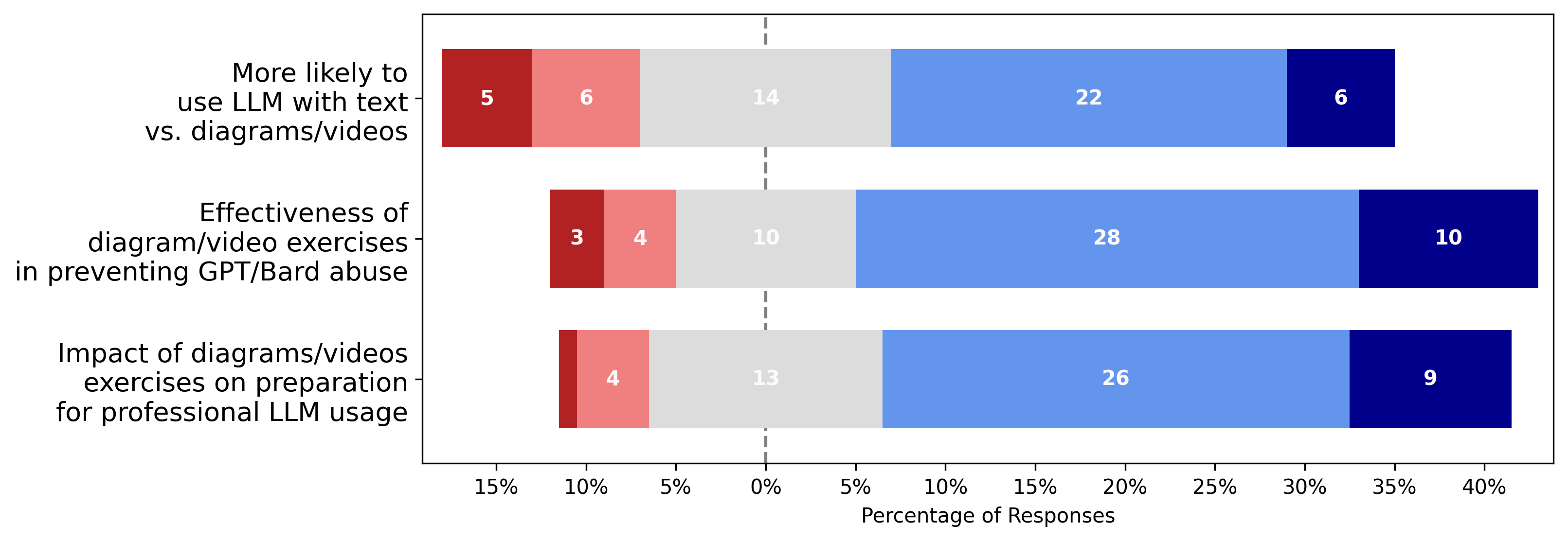}
  \vspace{-10pt}
  \caption{Student's opinions on diagram and video exercises' effect on their academic and professional LLM usage (RQ3).}
  \label{fig:table4}
\end{figure}

\subsection{RQ2: Interpretability and motivation of video-based projects}


Figure \ref{fig:table3} presents the results for each question related with RQ2. The vast majority of students indicated that videos make projects easier to interpret. However, with regards to the ease of note taking, most students feel that videos are not as good as textual descriptions.   Videos also seem to have a positive impact on students' motivation (Table \ref{tab:table_questions}: Question 3, RQ2). Finally, the majority of students reported that developing the project with the video statement is easier than with the traditional text-based assignments.


\subsection{RQ3: Use and Reliance on LLMs}


Figure \ref{fig:table4} presents results for questions related with RQ3. The majority of students indicate that they would be more likely to resort to an LLM when solving textual exercises than with diagrammatic or video exercises. The vast majority of students also agree that diagrams and videos are an effective way of preventing LLM abuse. Finally, students agree that these new formats will help them be more prepared for a potential professional future where they have to interact with an LLM to produce code ~\cite{alves2023centaur}.

\section{Discussion}

The proposed notation and diagrams present a new visual language relevant for the specificities of teaching and learning OO design and programming. These diagrams require that students first understand the problem by analysing the diagrammatic sample sets, and then proceed to define an object model and/or algorithm which solves it, with or without the help of a LLM.

Diagrams and videos can prescribe more or less directed tasks, according to educators' goals, by using clues such as `Implement in class Main', `Represent these concepts', and so on. However, educators should avoid exposing too much information textually or verbally, since some textual instructions can be parsed from images ~\cite{hou2023more} and audio interpretation tools are also emerging ~\cite{open-AI-chat-gpt-with-vision}.

We consider this model to have three advantages: (1) it forces students to interpret test cases and infer a problem description; (2), it should prevent direct `copy-and-prompting' from the assignment description to the LLM; and, (3) it requires students that wish to use an LLM to create prompts that guide it towards the goal. Although these last two ideas might seem contradictory, we believe that both have some pedagogical value, since it is important that students gain some experience with using LLMs in an authentic way, as helpers for solving coding problems, due to the likelihood that they will use these tools professionally ~\cite{alves2023centaur,barke2023grounded}.

During the experiment, we observed some interesting student behaviours which we believe should guide future work on this topic. One interesting aspect was, at least in some cases, students' difficulties in interpreting the diagrams prompted interesting discussions between students and teachers. When the doubts were not obvious, we would engage with the student and help them reach the expected interpretation by themselves. For example, when a student asked ``What does this `N + 1' tasks mean?'', the teacher pointed out that, before the function call, the object indicated `N tasks', and then asked the student what they thought was happening that resulted in the change from `N tasks' to `N + 1 tasks'. As for the videos, at least one student transcribed the video to a text document. We informally asked the student for their reasons, and they indicated doing it to support offline work because they didn’t have an internet connection at all times. One other student used a tool to download Youtube's automatic captions for the video assignment. Although these techniques could be used to exploit the video assignment, since not all information is obtainable from the narration, the videos should remain LLM-resistant for the time being.

Besides countering LLM over-reliance, the survey shows that these approaches also seem to have other benefits, such as the apparent positive impact on students' motivation levels. Moreover, the qualitative question also allowed us to identify another added benefit, since multiple students reported the need to analyse and reflect on the diagram's contents, in order to fully understand it. For example, participant S7 indicated that \textit{``Not knowing specifically what the function should do [...], I had to `spend' some time trying to understand that''}. Another participant, S39, commented that \textit{``Since I don't have a [function] name, I don't immediately know what I'm supposed to do, I have to think and reflect on the diagram''}. Finally, S56 indicated that \textit{``Sometimes, just the diagram without a brief contextual explanation made the exercise more difficult to understand than solving it.''}. As such, it appears that diagrams might also promote some development of analytical skills, and this would be an interest avenue to explore more rigorously in future work.

\subsection{GPT-4 and Bard's vision capabilities}




Recently, both GPT-4 and Bard have become capable of interpreting images. Given the potential threat to the proposed diagrams, we conducted initial explorations of the vision capabilities of these tools. To achieve this, GPT-4 and Bard where supplied with the same information that students had: a diagram containing some introductory text and Figure \ref{fig:diagram-task-state-transitions}, and then a second diagram which was equal to Figure \ref{fig:diagram-example-oop}. This experiment was repeated 3 times with the same images, but with slightly different prompts and the generated code was evaluated considering 3 compilation and 3 logical items. None of the attempts yielded a correct implementation of the `f06()' function, with both models generating code with compilation errors and/or logical errors \footnote{\textbf{GPT-4}'s diagram scores: Best compilation: 3/3 (experiment \#3). Best logic: 1/3 (experiment \#2). Worst compilation: 0/3 (experiments \#1 and \#2). Worst logic: 0/3 (experiments \#1 and \#3). \textbf{Bard}'s diagram scores: Best/Worst compilation: 0/3 (all experiments). Best logic: 1/3 (experiment \#2). Worst logic: 0/3 (experiments \#1 and \#3).}. In its best attempt, GPT-4 generated a function that compiled correctly, but the function's logic was hardcoded to the example provided in the diagram and would not work with different input values. Bard's best attempt was also problematic: it failed to create the function in the prescribed class, the argument order was wrong, it was not declared as final and the return-type was void instead of boolean. As for the logic, the code fails to correctly validate the `Task''s state transition and also fails to check if the `Employee' belongs to the company. The only sub-behaviour that would work as expected was the assignment of the Task to the Employee. Finally, instead of returning the requested boolean, an `Exception' was being thrown. We also supplied both models with an equivalent text-based description of the exercise. With the textual input, GPT-4 was able to generate an almost correct solution, failing only to declare the function as `static' \footnote{\textbf{GPT-4}'s textual scores: Compilation: 2/3. Logic: 3/3.}. Bard's solution to the text-based variant was also better than all its diagram-based attempts \footnote{\textbf{Bard}'s textual scores: Compilation: 3/3. Logic: 1.5/3.}. The logs of our experiments are available online\footnote{\url{https://doi.org/10.5281/zenodo.10547278}}.

This experiment shows that LLMs are much better at handling OOP exercises described textually, than diagram-based exercises. As such, diagrams seem a good approach to limit students' over-reliance on LLMs, at least for the time being.

\section{Limitations}






One limitation is the significant proportion of students opting for the neutral option in most quantitative questions. This could potentially have skewed some of our interpretations.


In relation to the questions about the impact of diagrams and videos on the utilization of LLMs, it is conceivable that some students may not have provided honest responses to this question given concerns around academic integrity. 

\section{Conclusions}


We believe that the biggest challenge in adapting courses and classes to this brave new world where LLMs with code generation capacity are easily accessible is creating exercises that are hard for an LLM to solve, but are still accessible and can be solved by students. In this paper we present a novel pedagogical approach to describe OO programming exercises and projects. This notation allowed us to replace the previously used natural language descriptions which could be handled, to some degree, by GPT-3.5, GPT-4 and Bard. The proposed notation was well received by our students, who also agree that it helped mitigate LLM over-reliance.

\bibliographystyle{splncs04}
\bibliography{gpt-vs-oop.bib}
%





\end{document}